\documentclass[sigconf, noacm]{acmart}
\renewcommand\footnotetextcopyrightpermission[1]{} 
\usepackage{listings}
\usepackage{subcaption}
\usepackage{tabularx}
\usepackage{multirow}
\usepackage{wrapfig}
\usepackage{etoolbox}
\makeatletter
\patchcmd{\maketitle}{\@copyrightspace}{}{}{}
\makeatother
\AtBeginDocument{%
  }

\definecolor{darkgreen}{RGB}{28,171,70}
\definecolor{darkred}{RGB}{182,21,14}

\newcolumntype{L}[1]{>{\raggedright\let\newline\\\arraybackslash\hspace{0pt}}m{#1}}
\newcolumntype{C}[1]{>{\centering\let\newline\\\arraybackslash\hspace{0pt}}m{#1}}
\newcolumntype{R}[1]{>{\raggedleft\let\newline\\\arraybackslash\hspace{0pt}}m{#1}}

\lstdefinestyle{pseudo}{
  frame=tb,
  language={c++},
  deletekeywords={with},
  aboveskip=2mm,
  belowskip=2mm,
  captionpos=b,
  showstringspaces=false,
  columns=flexible,
  basicstyle={\footnotesize\ttfamily},
  numbers=left,
  numberstyle=\tiny \color{black},
  keywordstyle=\color{blue},
  commentstyle=\color{magenta},
  frame=none,
  breaklines=true,
  breakatwhitespace=true,
  tabsize=3,
}

\setcopyright{none}

\newcommand{\smalltt}[1]{{\texttt{\small #1}}}

\begin{document}

\title{Towards Adaptive Storage Views in Virtual Memory}

\author{Felix Schuhknecht}
\affiliation{%
  \institution{Johannes Gutenberg University}
  \city{Mainz}
  \country{Germany}
}
\email{schuhknecht@uni-mainz.de}

\author{Justus Henneberg}
\affiliation{%
  \institution{Johannes Gutenberg University}
  \city{Mainz}
  \country{Germany}
}
\email{henneberg@uni-mainz.de}


\begin{abstract}

Traditionally, DBMSs separate their storage layer from their indexing layer. While the storage layer physically materializes the database and provides low-level access methods to it, the indexing layer on top enables a faster locating of searched-for entries. While this clearly separates concerns, it also adds a level of indirection to the already complex execution path.
In this work, we propose an alternative design: Instead of conservatively separating both layers, we naturally fuse them by integrating an adaptive coarse-granular indexing scheme directly into the storage layer. We do so by utilizing tools of the virtual memory management subsystem provided by the OS: On the lowest level, we materialize the database content in form of \textit{physical main memory}. On top of that, we allow the creation of arbitrarily many \textit{virtual memory storage views} that map to subsets of the database having certain properties of interest. This creation happens fully adaptively as a side-product of query processing. To speed up query answering, we route each query automatically to the most fitting virtual view(s). By this, we  naturally index the storage layer in its core and gradually improve the provided scan performance. 

\end{abstract}




\maketitle

\vspace*{-0.2cm}
\section{Introduction}
\vspace*{-0.1cm}
Classical DBMSs are separated into individual layers, where each layer serves a specific purpose. Two examples of this are the storage layer and the indexing layer. On the lowest level of the stack, the storage layer is responsible for physically materializing and maintaining the database. This includes providing low-level access methods to the individual records, such as \smalltt{getRecord(recordID)} or \smalltt{getRecordIterator()}. However, the storage layer does not have a notion of the semantics of the records, i.e., it cannot be asked to return records with a specific property. This is the responsibility of the indexing layer sitting on top of the storage layer. It maps properties, such as a specific value range, to a location in the store, where records with the property can be found. Consequently, it provides a high-level interface of the form \smalltt{getRecordsWithValue(keyRange)}, which translates the \smalltt{keyRange} to a list of qualifying \smalltt{recordID}s and utilizes \smalltt{getRecord(recordID)} of the storage layer to retrieve them. 

On the one hand, such a separation of concerns yields a clean system design, which is easy to maintain and to extend. However, on the other hand, introducing individual layers also comes at the cost of increasing the size and complexity of the system stack. This causes undesirable execution overhead by having to go through these layers during query processing.

In this work, we question whether strictly separating storage layer and indexing layer is reasonable at all, as both components are so tightly coupled by nature. We propose an alternative approach in the following: Instead of asking an indexing layer to point to the relevant parts of the database and to make the storage layer retrieve them, the storage layer should provide semantical (partial) \textit{views} on (subsets of) the database in the first place. Based on their predicates, all incoming queries are then routed only to the relevant view(s) in order to be answered, reducing the amount of data that need to be retrieved from the lowest layer of the stack already. 


\begin{figure}[ht!]
\vspace*{-0.2cm}
\includegraphics[page=2, width=.9\columnwidth, trim={3cm 4cm 1cm 3cm}, clip]{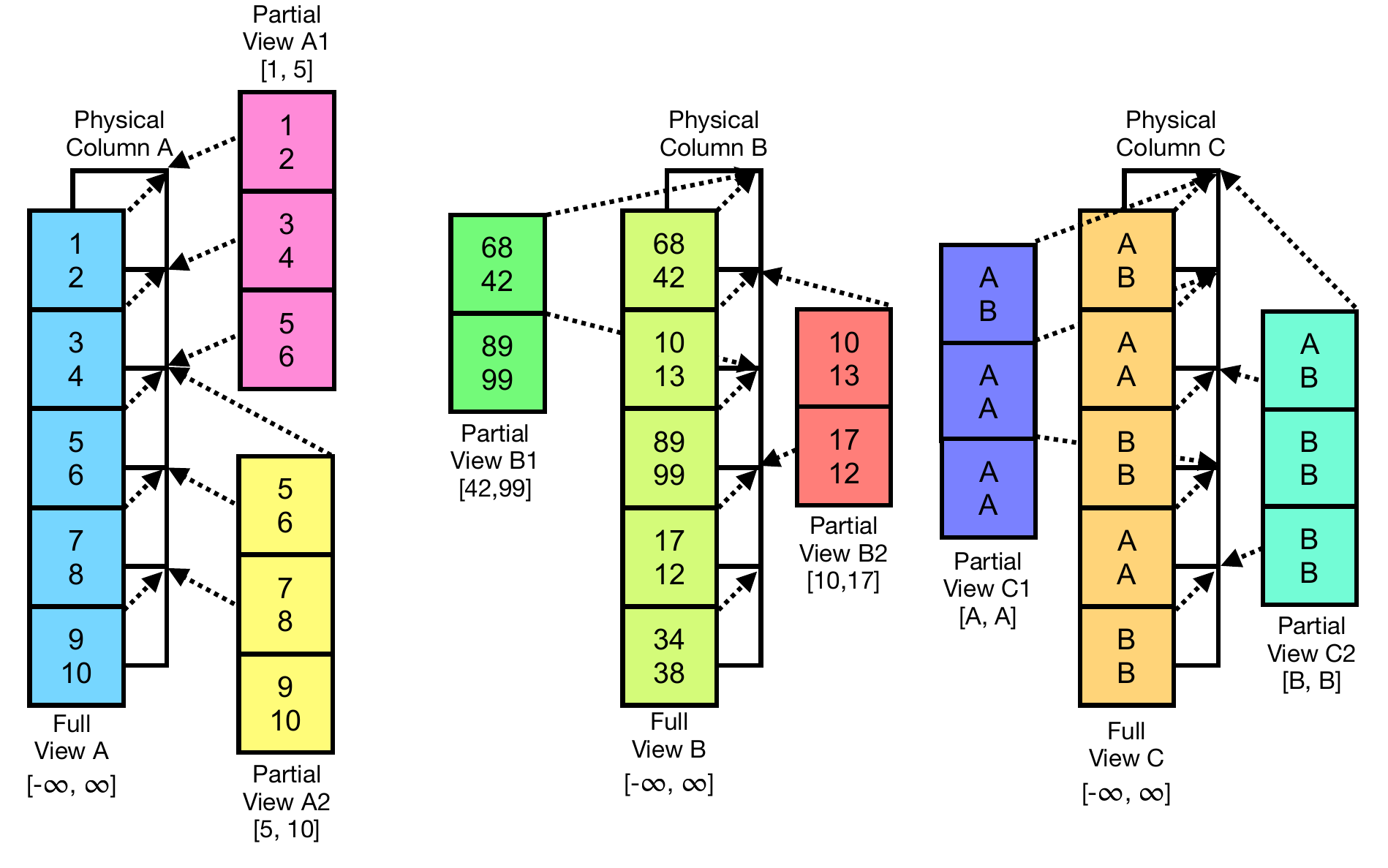}
\vspace*{-0.3cm}
\caption{The table representation in our adaptive storage layer. In addition to the full virtual view, each column provides two partial views indexing only subsets of the data.}
\label{fig:concept_new}
\vspace{-0.2cm}
\end{figure}

\vspace*{-0.2cm}
\subsection{Virtual Views}
\vspace*{-0.1cm}
Of course, such a solution could be engineered in software by integrating some sort of auxiliary coarse-granular index structure into the storage layer. 
However, this would just migrate the level of explicit indirection from the indexing layer to the storage layer.
As we target pure in-memory systems, we have a more sophisticated option available, which is strongly connected with how memory is represented in the system: By default, when allocating a memory area to hold our database, we actually allocate \textit{virtual} main memory that is internally mapped to \textit{physical} main memory by the OS. Thus, this virtual memory area resembles nothing but a \textit{view} on the physical memory underneath. By default, this virtual memory view, which is segmented into pages, spans the entire physical memory database.
However, this is not necessarily required to be the case: Using a technique called memory rewiring~\cite{rewiring}, it is possible to create virtual memory views that map only to a subset of the potentially scattered physical memory. If the underlying data is somehow clustered, this way of indexing can be very effective. Additionally, it is possible to update these virtual views freely at runtime, providing a large amount of flexibility, e.g., for reflecting updates. Also, multiple views can map to shared portions of physical memory, allowing us to create partially overlapping views.

Based on these observations, we \textbf{(1)~propose a storage layer design} as visualized in Figure~\ref{fig:concept_new} for a columnar layout. In addition to maintaining a \textit{full virtual view} denoted as ~$v_{[-\infty, \infty]}$, which covers the entire physical column, we allow the creation of multiple \textit{partial virtual views}~$v_{[l,u]}$. Each partial virtual view then indexes only the portion of the column that contains values within the range~$[l,u]$. 
As we will present in the following, we \textbf{(2)~create and maintain new partial views adaptively} and concurrently as a side-product of query processing. Given a set of views, we \textbf{(3)~route the queries to the most fitting view(s)} for query answering. Here we show a single-view mode, where exactly one view is used to answer the query, and a multi-view mode, where multiple overlapping views are considered. Also, we \textbf{(4)~discuss how existing partial views are updated efficiently} if the underlying physical database sees changes. Finally, we \textbf{(5)~perform an extensive experimental evaluation} that first evaluates the benefits of virtual indexing over explicit counterparts. Then, we evaluate the performance characteristics of our adaptive layer under four data distributions and show how scans can exploit the partial views over the query sequence. 

\vspace*{-0.2cm}
\subsection{Background: Memory Mappings}
\vspace*{-0.1cm}
Before diving into our adaptive storage layer, let us discuss the necessary background on manipulating virtual memory mappings. 

Traditionally, when allocating memory using an allocator like \smalltt{malloc()} or via \smalltt{mmap()}, we get \textit{virtual} main memory, which is internally mapped to \textit{physical} main memory by the OS. This physical memory is invisible to the programmer. While this is a convenient default behavior, it is undesired in our situation: To actively create our views, we need to acquire handles on both virtual and physical memory and actively manipulate the mapping between these two types of memory during runtime. Fortunately, a technique that grants this possibility is memory rewiring~\cite{rewiring, anyolap}. The core idea is to introduce physical memory to user-space in the form of \textit{main-memory files}. A main-memory file acts like a normal file, but it is backed by volatile (physical) main memory pages, instead of disk pages. Thus, a main-memory file can serve as a handle to physical memory.   
By creating a virtual memory area that maps to such a main-memory file using \smalltt{mmap()}, we can establish a controllable mapping from virtual to physical memory. A nice side effect of this approach is that we can update the mapping freely at page granularity during runtime. Note that memory rewiring is fully supported by the vanilla Linux kernel and that utilizing the techniques adds only a negligible overhead for the very first page access after (re-)mapping~\cite{rewiring}.   

%
%
%
%

\vspace*{-0.2cm}
\section{Adaptive Storage Layer}
\vspace*{-0.1cm}
At the core, our adaptive layer maintains for each column of each table the following components: (a)~A physical column materializing all contained values. (b)~A set of virtual views~$V=\{v_{[-\infty,\infty]}, v_{[l_0, u_0]}, v_{[l_1, u_1]}, ...\}$ that map to subsets of the physical column. By default, only a full view $v_{[-\infty,\infty]}$ exists, but further partial views are created on demand as we will describe in Section~\ref{ssec:adaptive_views}. Apart from the virtual memory address, for each virtual view~$v_{[l_i, u_i]}$, we only materialize the covered value range $[l_i, u_i]$ and its size in number of pages. 

The creation of a new partial view covering the value range~$[l,u]$ always happens based on one or multiple existing views that fully cover $[l,u]$.  
To perform the creation, we first allocate a fresh virtual memory area representing~$v_{[l,u]}$ using \smalltt{mmap()}, which we back by anonymous memory. This first call to \smalltt{mmap()} acts as a mere reservation of virtual memory for our view and is almost for free. In this step, we over-allocate the memory area to the size of the entire column, as we are unaware of how many physical pages will qualify and thus be mapped by the view. Now, we scan and filter the overlapping existing view(s) for distinct qualifying physical pages with respect to $[l, u]$. When we find a qualifying page, we remap a currently anonymously mapped virtual page of $v_{[l,u]}$ to the physical page via a call to \smalltt{mmap()} using the \smalltt{MAP\_FIXED} flag. Eventually, all views have been fully scanned and the new view $v_{[l,u]}$ maps to all qualifying physical pages. 
To be able to use the views during query processing, we need to store a small amount of meta-data: As partial views might map to arbitrary subsets of the physical column, we have to embed an $8$B~pageID at the beginning of each physical page. When scanning a partial view, this pageID is used to identify for each read value to which tuple it belongs.

\vspace*{-0.2cm}
\subsection{Query Routing}
\label{ssec:query_routing}
\vspace*{-0.1cm}

To answer an incoming query using the existing views, we support two modes of operation:

In \textit{single-view} mode, we use exactly one view to answer the query, where this view must fully cover the predicates of the query. If there are multiple views available that fulfill this property, we pick the view that indexes the smallest amount of physical pages to minimize the scanning effort. 

In \textit{multi-view} mode, we potentially use multiple views to answer a single query, provided that these multiple views fully cover the requested range in conjunction. As physical pages might be shared between multiple partial views, we additionally have to keep track of processed physical pages to avoid scanning a page twice, as this would lead to incorrect results. We realize this using a fixed-size bitvector. Currently, when running this mode, the system tries to answer a query using multiple views if possible, instead of directing the query to a single (potentially larger) view. In future work, we plan to base this decision on the covered value ranges and the number of indexed pages.   

\vspace*{-0.2cm}
\subsection{Adaptive Partial Views}
\label{ssec:adaptive_views}
\vspace*{-0.1cm}

Our storage layer creates and maintains partial views \textit{adaptively} and \textit{transparently} as a side-product of query processing. In Listing~\ref{listing:adaptive_partial_views}, we present the pseudo-code for the partial view creation. 

Given a query~$q$ selecting range $[l_q,u_q]$, in line~\ref{listing:start_get_optimal_view}, the system first retrieves the most fitting set of existing views that fully cover the range $[l_q,u_q]$. To avoid the re-processing of shared physical pages in multi-view mode, we keep track of the already traversed pages in a bitvector, as described in Section~\ref{ssec:query_routing}. 
In lines~\ref{listing:start_new_partial_view}-\ref{listing:end_new_partial_view}, as a side-product of query answering, we create a new partial view on the column covering the value range $[l_q, u_q]$. However, this partial view can turn out to cover even more values than $[l_q, u_q]$. To find out, we maintain the largest value~$l'_q < l_q$ as well as the smallest value $u'_q > u_q$ that we observe over all non-qualifying pages. Consequently, all values strictly between $l'_q$ and $u'_q$ \textit{must} be stored on qualifying pages. Therefore, we are allowed to extend the covered range of our new partial view from $[l_q, u_q]$ to $[l'_q + 1, u'_q - 1]$.

Next, in lines~\ref{listing:start_suggest_partial_view}-\ref{listing:end_suggest_partial_view}, the system decides whether the new partial view~$v_{[l'_q + 1, u'_q - 1]}$ will be retained or whether it should be discarded. To decide this, we first test whether the new partial view indexes less physical pages than the full view. If so, then this view improves over the full view and should be considered further.  
The next question is how the new partial view relates to existing partial views.  
We first check whether the new partial view covers only a \textit{subset} of an existing partial view, but references a similar amount of physical pages. In this case, we also want to reject the new partial view, as it covers a smaller value range and is thus less useful than the existing partial view, but causes a similar amount of work during query answering. When comparing the views in terms of number of indexed physical pages, a discard tolerance~$d$ is taken into account, which can be set by the user. A new partial view covering a subset of an existing partial view is then discarded, even if it indexes $d$ physical pages less than the existing view.  
We also test whether the new partial view covers a \textit{superset} of an existing partial view and is thus a candidate for replacing that view. If the new partial view has a size similar to the one to replace, then we consider it more useful and replace the old view by it. Again, a replacement tolerance~$r$ is taken into account, where replacement is happening only if the new partial view indexes at most $r$ physical pages more than the existing partial view.
\begin{lstlisting}[style=pseudo, caption={Pseudo-code of adaptive partial view maintenance.}, label={listing:adaptive_partial_views}, xleftmargin=4.0ex, escapechar=|]
answerQueryAndMaintainViews(q):
  // get optimal view
  |\label{listing:start_get_optimal_view}|optimalViews = views.getOptimalViews(q.l, q.u) |\label{listing:end_get_optimal_view}|
  lNew, uNew = optimalViews.coveredValueRange()
  // create new partial view while answering query |\label{listing:start_new_partial_view}|
  queryResult = {}
  candidateView = createEmptyPartialView(q.l, q.u)
  processedPages = createEmptyBitvector(fullView.numPages())
  for(view: optimalViews)
    for(page: view.pages())
      if(processedPages[page.getPageID()] == 0)
        (pageResult, minValue, maxValue) = page.scanAndFilter(q)
        if(pageResult == {})
          lNew = max(lNew, minValue+1)
          uNew = min(uNew, maxValue-1)
        else
          queryResult += pageResult
          candidateView.addPage(page)
        processedPages[page.getPageID()] = 1
  candidateView.updateRange(lNew, uNew) |\label{listing:end_new_partial_view}|
  // suggest partial view to view index  |\label{listing:start_suggest_partial_view}|
  if(candidateView.numPages() < fullView.numPages())
    for(partialView: views.getPartialViews())
      if(candidateView.coversSubsetOf(partialView) &&
         candidateView.numPages() >= partialView.numPages() - d)
        delete(candidateView)
        return queryResult
      if(candidateView.coversSupersetOf(partialView) &&
         candidateView.numPages() <= partialView.numPages() + r)
        views.replace(partialView, candidateView)
        return queryResult
    views.insert(candidateView) |\label{listing:end_suggest_partial_view}|
  return queryResult
\end{lstlisting}
If the new partial view does not relate to an existing partial view regarding the above criteria, we include it into our set of views, but only if the maximum number of views has not been reached yet. If the limit has been reached already, we stop the generation of new partial views altogether and perform query answering based on the static set of existing views. 

\subsection{Optimized View Creation}
\label{ssec:optimizations}

As repetitive calls to \smalltt{mmap()} are the most expensive part of view creation, we perform two optimizations when creating new views: 

(1)~We map consecutive qualifying physical pages in a single call. While scanning the existing views, we keep track of consecutive physical pages that qualify. As soon as we encounter a non-qualifying page, we map all previously seen qualifying pages in one call. This minimizes the amount of required \smalltt{mmap()} calls.

(2)~We perform the actual \smalltt{mmap()} calls in a separate thread. Instead of letting the scanning thread map each qualifying page, it only inserts a request to map the physical page into a concurrent queue from the Boost library. A separate mapping thread constantly polls this queue and performs the actual \smalltt{mmap()} calls. When the new partial view is completely mapped, the mapping thread informs the main thread that it can be inserted into the view index as it is ready for the upcoming query processing.

\vspace*{-0.2cm}
\subsection{Handling Updates}
\label{ssec:updates}
\vspace*{-0.1cm}

If updates happen through the full views, these updates must be reflected by all existing partial views to ensure correctness. This involves potentially adding and removing pages from each partial view that covers a value range affected by an update. As this process can become costly when being performed for each update individually, we support the adjustment of partial views with respect to an adjustable batch of updates. In the following, we outline the steps necessary to align a single partial view covering the value range $[a,b]$. We assume that the indexed column has seen a sequence of $n$ updates of the form $U=[(r_0, old_0, new_0),...,(r_{n-1}, old_{n-1}, new_{n-1})]$. Here, $r_i$~describes the row written to, $old_i$ is the old value overwritten by the update, and $new_i$ represents the new value written. 

In the first step, we filter the sequence of updates~$U$ such that only the very last update to each row remains reflected. Precisely, if the batch contains updates of the form $u_0=(r_a, old_i, new_i)$, $u_1=(r_a, old_j, new_j)$, $u_2=(r_a, old_k, new_k)$, where first $u_0$, then $u_1$, and then $u_2$ are applied, then we replace these three updates with a single update of the form $(r_a, old_i, new_k)$ to reflect only the original value as well as the last written value. This results in a new sequence of updates~$U'$.
In the second step, we group all updates of $U'$ by each modified physical page and test for each page~$p$, whether it is already indexed by this partial view. We have to differentiate between two situations: (1)~$p$ is not indexed, but should be indexed due to the updates. (2) $p$ is currently indexed, but should not be indexed anymore.

In case~(1), we check whether at least one update~$u_i$ of $p$ contains a new value $new_i \in [a,b]$. If it has seen such an update, we map an "unused" virtual page of the partial view to the physical page to index it. Otherwise, we do not change the partial view in any way. Remember that we have "unused" virtual pages available, as we perform an over-allocation during view creation.
In case~(2), more properties need to be considered to make the decision. If no new value $new_i$ of any update~$u_i$ falls into the value range of the partial view, i.e., $new_i \notin [a,b]$, there might still be other values of that page that fall into the range. Therefore, we first check whether at least one old value $old_i$ was covered by this partial view, i.e., $old_i \in [a,b]$. If this is not the case, then none of the updates on this physical page affect this partial view in any way and it can remain indexed (as clearly at least one other value on that page is part of $[a,b]$ -- otherwise, it would not be indexed). However, if at least one old value $old_i \in [a,b]$, we have to inspect all values of the physical page. Only if none of these values fall into $[a,b]$, we are allowed to remove it from the index. 

\vspace*{-0.2cm}
\subsection{Querying Memory Mappings}
\label{ssec:mapping_status}
\vspace*{-0.1cm}

In order to update our partial views as described in Section~\ref{ssec:updates},
we need to obtain the current mapping between virtual pages and physical pages.
Fortunately, the Linux kernel exposes all memory mappings through the \smalltt{/proc} virtual filesystem. For each process with ID \smalltt{PID}, mapping information is listed in the virtual file \smalltt{/proc/PID/maps}. The file has the format

\noindent
\begin{minipage}{\linewidth}
\begin{footnotesize}
\begin{lstlisting}
address           perms offset  dev   inode   pathname
08048000-08056000 rw-s 00002000 03:0c 64593   /dev/shm/db
\end{lstlisting}
\end{footnotesize}
\end{minipage}

\noindent where the first three columns contain the mapped virtual address range (\smalltt{address}), the permissions (\smalltt{perms}), and the offset into the main memory file (\smalltt{offset}). 
As parsing this file is costly if a sufficient amount of mappings exist, we do not want to perform it frequently. Instead, we parse the file only once before applying a batch of updates. We materialize the parsed mappings page-wise in a bi-directional map (Boost bimap), which is maintained from user-space during the update process. After the batch of updates has been processed, we can safely discard the bimap again.

\vspace*{-0.2cm}
\section{Experimental Evaluation}
\label{sec:experiments}
\vspace*{-0.1cm}

In the following, we first first justify our design based on a micro-benchmark (Section~\ref{ssec:indirection_variants}). Then, we evaluate the behavior and performance charactertistics of the adaptive storage layer (Section~\ref{ssec:eval_query_processing}, \ref{ssec:eval_optimizations}, and \ref{ssec:eval_updates}).
We perform all experiments on an Intel Core i7~12700KF @ 5GHz with 64GB of DDR5-4800 RAM, where we activate only the eight performance cores in the bios. The OS is a 64-bit Ubuntu 22.04 LTS with a vanilla Linux kernel in version~5.15. Note that our code requires a \smalltt{tmpfs}~\cite{tmpfs} filesystem being mounted, which is by default the case under \smalltt{/dev/shm/} in the case of Ubuntu. Our adaptive layer purely operates with $4$KB small pages.
No root permissions are required to execute our code. However, we increase the amount of allowed memory mappings from the default of $2^{16}-1$ to $2^{32}-1$.
Discard and replacement tolerance are both set to $0$ in all experiments. Apart from that, in all experiments, we report the average time of three runs.

\begin{figure}[h!]
  \centering
    \begin{subfigure}[b]{.32\columnwidth}
        \centering
	\includegraphics[width=\textwidth, trim={0 0 9cm 0}, clip]{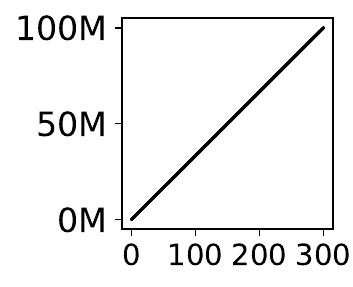}
    \caption{Linear dist.}
    \label{fig:linear_dist}
  \end{subfigure}
  \hspace*{0.1cm}
  \begin{subfigure}[b]{.32\columnwidth}
    \centering
	\includegraphics[width=.75\textwidth, trim={10.5cm 0 0 0}, clip]{data.pdf}
    \caption{Sine dist.}
    \label{fig:sine_dist}
  \end{subfigure}
    \hspace*{-0.2cm}
     \begin{subfigure}[b]{.32\columnwidth}
         \centering
	\includegraphics[width=.75\textwidth, trim={6cm 0 4.5cm 0}, clip]{data.pdf}
    \caption{Sparse dist.}
    \label{fig:sparse_dist}
  \end{subfigure}
   \vspace*{-0.2cm}
  \caption{Clustered data distributions.}
  \label{fig:distributions}
  \vspace*{-0.2cm}
\end{figure}

Apart from a uniform distribution, we run experiments on the three data distributions shown in Figure~\ref{fig:distributions}, reflecting clustered data distributions, as seen in time series or sensor data. The $x$~axis shows the pageID and the $y$~axis shows the generated values. The sine distribution cycles every 100 pages, whereas for the sparse distribution, 90\% of the pages are filled with zeros. 

\vspace*{-0.2cm}
\subsection{Partial Views: Explicit vs Virtual}
\label{ssec:indirection_variants}
\vspace*{-0.1cm}

We start by experimentally comparing the query performance of a partial view, where the qualifying pages are indexed explicitly, to that of a virtual partial view. For the explicit partial view, we test three possible variants:
Variant ``Zone Map'' stores the observed minimum and maximum value of each page in-place at the beginning of the page, before the actual values are materialized. During a scan, non-qualifying pages are simply skipped. Variant ``Bitmap'' maintains a separate bitvector, in which a one denotes that a page qualifies. A lookup basically results in a scan of the bitvector with subsequent jumps into the column for each qualifying page. Variant ``Vector of Page-IDs'' maintains a vector containing only IDs of qualifying pages. A lookup utilizes the IDs to locate the actual pages in the column. Note that this variant can benefit from prefetching to speed up lookups to subsequent pages. Thus, when starting to process the page at address \smalltt{pages[i]}, we already advise to prefetch the next page at address \smalltt{pages[i+1]} using the GCC intrinsic \smalltt{\_\_builtin\_prefetch(pages[i+1], 0, 0)}. 
Also, we include the Variant ``Physical Scan'', which resembles scanning a consecutive memory area, that has been allocated traditionally with \smalltt{new} and already contains all qualifying pages. This resembles an artificial optimal baseline.

\begin{figure}[h!]
	\vspace*{-0.2cm}
	\includegraphics[width=.85\linewidth, trim={0 0 0 0}, clip]{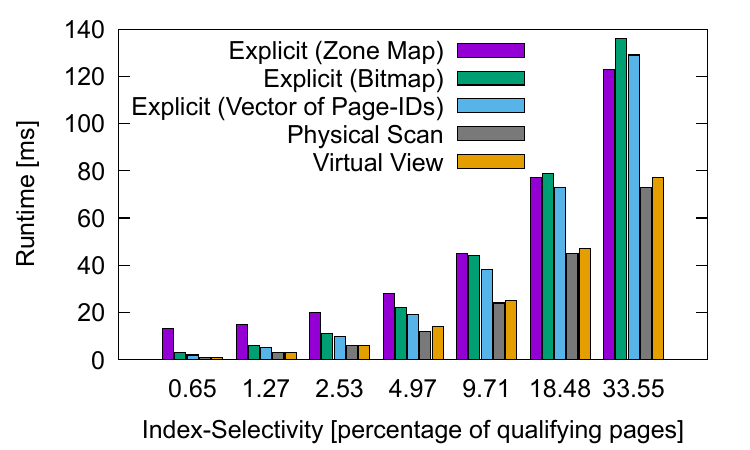}
		\vspace*{-0.2cm}
	\caption{Query performance of explicit vs virtual views.}
	\label{fig:partial_view_filter}
	\vspace*{-0.2cm}
\end{figure}

\begin{figure*}[h!]
  \centering
  \begin{subfigure}[b]{.33\textwidth}
	\includegraphics[width=\linewidth, trim={0.2cm 0 0.7cm 0}, clip]{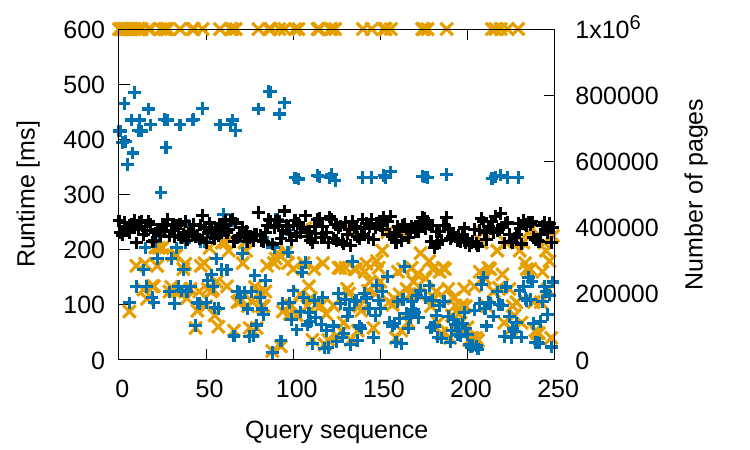}
    \caption{Sine distribution.}
    \label{fig:query_processing_single_view_sine_shuffled}
  \end{subfigure}
  \begin{subfigure}[b]{.33\textwidth}
	\includegraphics[width=\linewidth, trim={0.2cm 0 0.7cm 0}, clip]{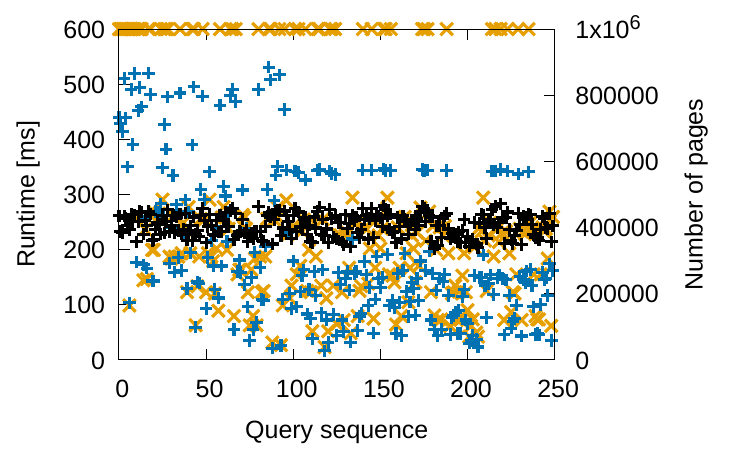}
    \caption{Linear distribution.}
    \label{fig:query_processing_single_view_linear_shuffled}
  \end{subfigure}
     \begin{subfigure}[b]{.33\textwidth}
	\includegraphics[width=\linewidth, trim={0.2cm 0 0.7cm 0}, clip]{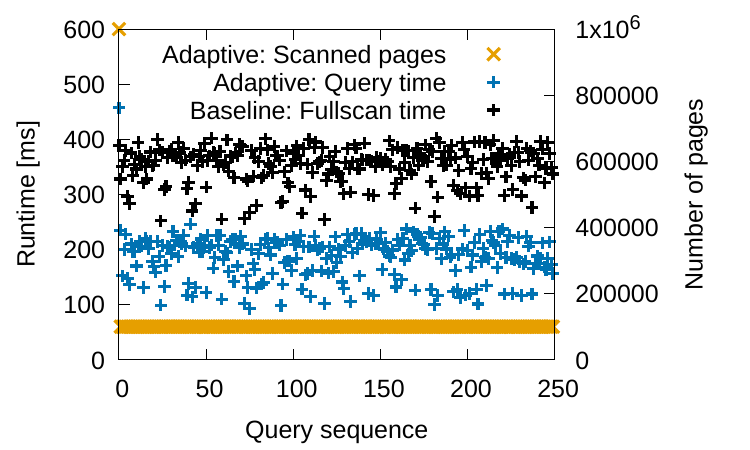}
    \caption{Sparse distribution.}
    \label{fig:query_processing_single_view_sparse_shuffled}
  \end{subfigure}
  \vspace*{-0.5cm}
  \caption{Adaptive query processing using single-view mode.}
  \label{fig:query_processing_single_view}
    \vspace*{-0.3cm}
\end{figure*}

In the experiment, we allocate a column consisting of $1$M pages of size $4$KB filled with uniformly generated random $8$B integers in the range of $[0, 100\text{M}]$. We now create a single partial view that indexes all pages containing values in the range of $[0, k]$, where we vary $k$ in logarithmic steps to simulate different index selectivities from $k=1{,}250$, indexing $0.65\%$ of all pages, up to $k=80{,}000$, indexing $33.55\%$ of all pages. 
After creating the partial view, we also update $10{,}000$ uniformly selected entries to simulate a change of the partial view. This can impact the query performance, as depending on the variant, the updates might scatter the order in which pages are indexed. Finally, for each $k$, we then answer a query selecting the range $[0, k/2]$, to select only $50\%$ of the data, and report the runtime.
Figure~\ref{fig:partial_view_filter} shows the results for all four variants. Using zone maps to explicitly index the pages is in all cases the most expensive option, as the meta-data of all pages must be inspected, involving $1$M~address translations. The bitmap and the vector approach perform slightly better. In all cases, virtual partial views clearly win, as it has the least code complexity and naturally exploits hardware prefetching. 

\vspace*{-0.2cm}
\subsection{Adaptive Query Processing}
\label{ssec:eval_query_processing}
\vspace*{-0.1cm}

Let us now see how our adaptive storage layer performs query processing and how it dynamically creates and utilizes partial views. 

In Figure~\ref{fig:query_processing_single_view}, we start with an evaluation of the single-view mode. We allow the system to create up to $100$~views adaptively and test three clustered distributions, namely sine, linear, and sparse, as visualized at the beginning of Section~\ref{sec:experiments} on a single-column table of $1$M pages.  We generate a sequence of $250$~queries which vary the selected value range step-wise from $50$M (low selectivity) down to $5000$ (high selectivity). Before firing, we shuffle the generated queries randomly. Aside from the response time of each query, we report the number of scanned physical pages, as it shows whether and how the views are utilized. As a baseline, we plot the response time when only full scans of the whole column are used to answer the queries.
We can see that our adaptive partial view creation indeed speeds up query processing significantly in all tested cases. However, as expected, it takes a couple of queries until a sufficient amount of partial views have been created that cover the value ranges of incoming queries. In the early phase of the sequence (up to around 50 for sine and linear), most queries are answered with a full scan, and creating a new view alongside adds some overhead. However, this work pays off during later phases of the query sequence, where a large amount of queries can be answered from a partial view. This also becomes visible when inspecting the number of scanned pages.

\begin{figure}[ht!]
  \centering
    \begin{subfigure}[b]{.50\columnwidth}
	\includegraphics[width=\linewidth, trim={0.2cm 0 0.55cm 0}, clip]{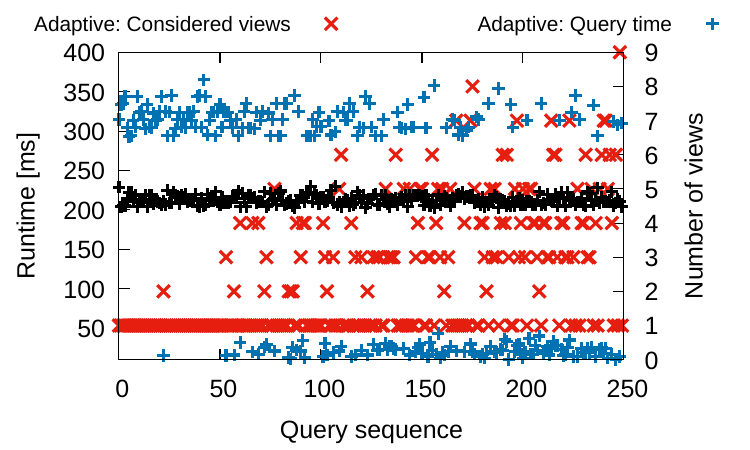}
    \caption{Sine distribution (sel.~1\%).}
    \label{fig:query_processing_multi_view_sine_1M}
  \end{subfigure}
  \hspace*{-0.2cm}
  \begin{subfigure}[b]{.50\columnwidth}
	\includegraphics[width=\linewidth, trim={0.2cm 0 0.55cm 0}, clip]{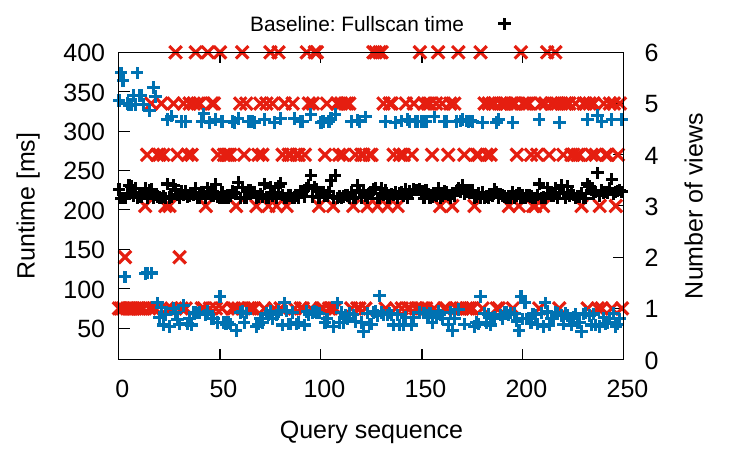}
    \caption{Sine distribution (sel.~10\%).}
    \label{fig:query_processing_multi_view_sine_10M}
  \end{subfigure}
    \vspace*{-0.5cm}
  \caption{Adaptive query processing using multi-view mode.}
  \label{fig:query_processing_multi_view}
    \vspace*{-0.3cm}
\end{figure}

In Figure~\ref{fig:query_processing_multi_view}, we evaluate the multi-view mode, where multiple views are used to answer a query, if they fully cover the requested value range. We see this as the mode of choice for queries with a fixed selectivity, as it highly increases the chance of partial view usage. Thus, in this experiment, we fix the selectivity and focus on the previously used sine distribution. We test both a selectivity of $1\%$ and of $10\%$. As with a higher selectivity, it takes longer to cover the column with partial views, we allow the creation of up to $200$~views for $1\%$~selectivity. For $10\%$, we limit the number of views to $20$. Apart from the runtime, we also show the number of used views per query. 
From the results we can see that the multi-view mode is indeed utilized by the query processing. For $1\%$ selectivity, up to 9 views are considered, for $10\%$, up to 6 views. We can see that if multiple partial views can be considered, the performance drastically improves over a full scan. 

In Table~\ref{tab:accumulated_query_processing}, we also show the accumulated query response time over the entire sequence, which confirms our impression: The adaptive variant improves over full scans by up to a factor of $1.88$x.

\begin{table}[h!]
\vspace*{-0.2cm}
\small
\setlength\tabcolsep{0.10cm}
\begin{tabular}{r || C{0.9cm} | C{0.8cm} | C{0.8cm} || C{0.8cm} | C{0.9cm}}
 Mode & Fig.~\ref{fig:query_processing_single_view_sine_shuffled} & Fig.~\ref{fig:query_processing_single_view_linear_shuffled} & Fig.~\ref{fig:query_processing_single_view_sparse_shuffled} & Fig.~\ref{fig:query_processing_multi_view_sine_1M} & Fig.~\ref{fig:query_processing_multi_view_sine_10M}\\\hline
 Full scans only & 58.6s & 60.9s & 88.2s & 53.2s & 55.2s\\
 Adaptive view selection& 41.2s & 49.4s & 46.7s & 46.0s & 35.8s\\
\end{tabular}
\caption{Accumulated response time over all 250 queries.}
\vspace*{-0.6cm}
\label{tab:accumulated_query_processing}
\end{table}

\vspace*{-0.2cm}
\subsection{Impact of Optimizations on View Creation}
\label{ssec:eval_optimizations}
\vspace*{-0.1cm}

\begin{figure}[h!]
  \vspace*{-0.2cm}
  \centering
  \begin{subfigure}[b]{.49\columnwidth}
	\includegraphics[width=\linewidth, trim={0.45cm 0cm 5.35cm 4cm}, clip]{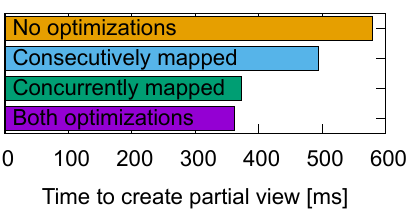}
    \caption{Uniform distribution.}
    \label{fig:optimizations_uniform}
  \end{subfigure}
  \begin{subfigure}[b]{.49\columnwidth}
	\includegraphics[width=\linewidth, trim={0.45cm 0cm 5.35cm 4cm}, clip]{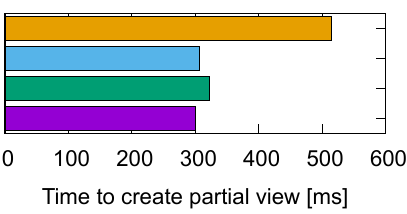}
    \caption{Sine distribution.}
    \label{fig:optimizations_sine}
  \end{subfigure}
    \vspace*{-0.2cm}
  \caption{Impact of optimizations on view creation.}
  \label{fig:optimizations}
    \vspace*{-0.2cm}
\end{figure}

Let us now evaluate the impact of the applied optimizations for our partial view creation. In the experiment of Figure~\ref{fig:optimizations}, we compare the time to create a single partial view on a column of $3.9$GB (a)~without optimizations, (b)~when mapping consecutive qualifying physical pages in one \smalltt{mmap()}, (c)~when mapping in a separate thread, and (d)~with both optimizations activated. In Figure~\ref{fig:optimizations_uniform}, we create a partial view~$v_{[0, 100\text{k}]}$ on a uniform distribution from $[0,100\text{M}]$, indexing almost $400\text{k}$~pages. In Figure~\ref{fig:optimizations_sine}, we create a view on $v_{[0, 2^{63}]}$ on a sine distribution from $[0,2^{64}-1]$, indexing around $520\text{k}$~pages.
We can clearly observe a significant impact of both optimizations. The effect of each individual optimization depends on the underlying distribution: If a clustering is present in the data, such as for the sine distribution, more pages can be mapped in one \smalltt{mmap()} call. Consequently, this optimizations pays of significantly. In comparison, mapping concurrently in a background thread is more independent from the distribution. In total, the optimizations improve the performance by a factor of $1.6$x (uniform) to $1.7$x (sine).

\vspace*{-0.2cm}
\subsection{Update Performance}
\label{ssec:eval_updates}
\vspace*{-0.1cm}

\begin{figure}[h!]
  \vspace*{-0.2cm}
  \centering
  \begin{subfigure}[b]{.49\columnwidth}
	\includegraphics[width=\linewidth, trim={0.2cm 0 0.7cm 0}, clip]{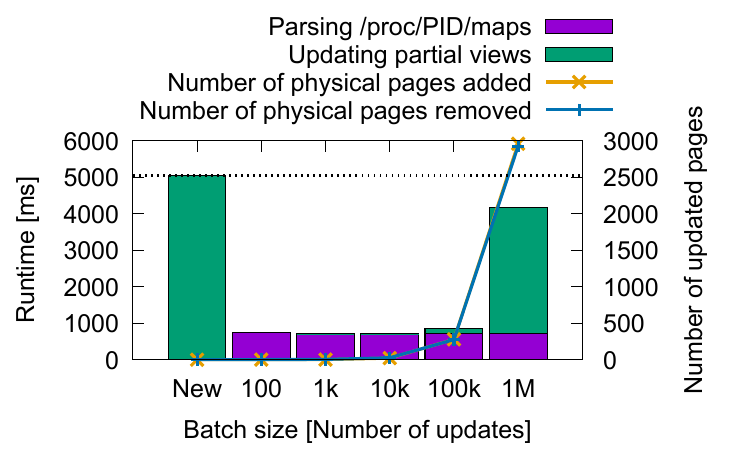}
    \caption{Uniform distribution.}
    \label{fig:updates_5views_uniform}
  \end{subfigure}
  \begin{subfigure}[b]{.49\columnwidth}
	\includegraphics[width=\linewidth, trim={0.2cm 0 0.7cm 0}, clip]{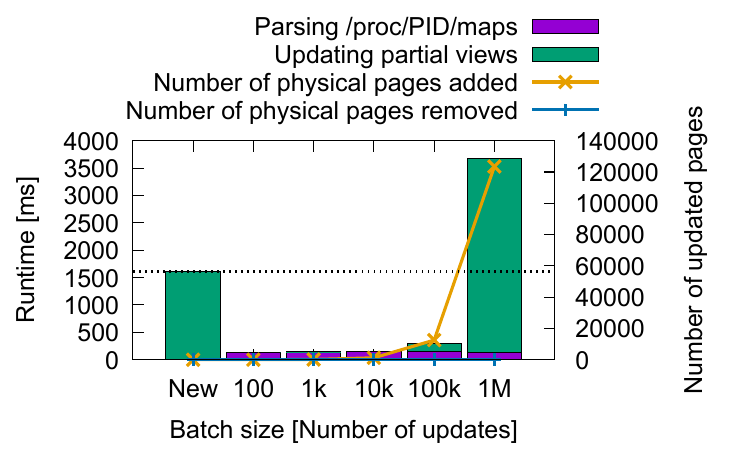}
    \caption{Sine distribution.}
    \label{fig:updates_5views_sine}
  \end{subfigure}
    \vspace*{-0.2cm}
  \caption{Update performance when varying the number of updates applied in a batch.}
  \label{fig:updates}
    \vspace*{-0.3cm}
\end{figure}

Regarding Figure~\ref{fig:updates}, we now evaluate the time it takes to update a set of partial views if changes happen to the underlying table. To set up the experiment, we first create a table 
with one column consisting of $1$M~pages. In the first experiment (Figure~\ref{fig:updates_5views_uniform}), we fill the column with uniformly distributed integers within $[0, 2^{64}-1]$. In the second experiment (Figure~\ref{fig:updates_5views_sine}), we test a sine distribution of the same range. In both cases, we create five partial views on the column, where each view covers a randomly selected $1/1024\text{-th}$ of the value range of the column. This results in indexing around $400$k~physical pages for the uniform data and in $20$k~pages for the sine case.
We now perform a varying number of updates ($100$ to $1$M in logarithmic steps) and update all existing partial views accordingly. We split the total time into the time required to parse the mapping file (as described in Section~\ref{ssec:mapping_status}) and the time required to update the partial views (as described in Section~\ref{ssec:updates}). Additionally, we show the time to (re-)build all five views from scratch instead of applying our updating algorithm. On a second $y$~axis, we also plot the number of physical pages added/removed during the update process, as this relates to the measured runtime.

From the results, we can learn that updating existing views is indeed more beneficial than re-building all views in almost all cases. Only for a very large batch size of $1$M, a rebuild pays off for the sine distribution. Further, we see that the cost of parsing is visible (and dominant) if the batch size is small. We also see that parsing is more costly under a uniform distribution than under the sine distribution. This is due to the fact that the clustered sine distribution results in partial views indexing more physical pages consecutively, resulting in less memory mappings and a smaller \smalltt{maps} file to parse. Finally, we learn that updating the partial views remains cheap unless a very large number of updates must be applied. From the comparison of both plots, we can see that removing pages is considerably more expensive than adding pages, as removing a page might require a full page scan. 

\vspace*{-0.2cm}
\section{Related Work}
\vspace*{-0.1cm}

This paper builds upon other work that actively utilizes virtual memory features to enhance database components: In~\cite{rewiring, phd_thesis}, the authors utilize the manipulation of memory mappings to speed up data structures (vector) and algorithms (partitioning). In~\cite{rewired_cracking, webassembly, pma}, the authors generally exploit the technique to avoid physical copying whenever possible.
Also, virtual memory features have been exploited actively in the context of snapshotting. The works~\cite{rewiring, anyolap} provide an in-process solution to the problem while \cite{hyper} snapshots virtually by spawning processes using the system call \smalltt{fork()}.
Apart from the memory perspective, there exists work that introduces views to the storage level in the context of the "one size fits all" movement. E.g., in OctopusDB~\cite{octopusdb, octopusdb2}, different storage views can be adaptively created to represent a logical database. While being related to our layer, a core difference to our work is that OctopusDB creates physical and potentially redundant representations of the database, while we purely create virtual views that do not incur redundancy of physical data.  
From the indexing perspective, our work also relates to adaptive indexing~\cite{uncracked_pieces, cracking1, cracking2, cracking3, cracking4}. Therein, the indexing state is gradually refined as a side-produce of query processing, however, the reorganization happens on the physical level as well. Also, coarse-granular indexing, in particular zone-maps~\cite{zone_maps}, where value ranges of pages are materialized and utilized during query processing, share similarities with our partial views. As in~\cite{time_series1, time_series2}, our technique is most effective for clustered data, such as time series of e.g., sensor data. 

\vspace*{-0.2cm}
\section{Conclusion}
\vspace*{-0.1cm}

In this work, we discussed how adaptive storage views can be expressed in virtual memory to naturally embed coarse-granular indexing in the storage layer. We showed an optimized adaptive strategy to construct partial views as a side-product of query processing and discussed how to handle updates efficiently. In our experimental evaluation, we showed that virtual views offer a better query performance than explicit alternatives and that our adaptive storage layer gradually improves query processing performance over full scans by up to a factor of 1.88x on clustered data. 

\noindent 
\textbf{Material:} All code, material, and results of this paper are available under \href{https://gitlab.rlp.net/fschuhkn/adaptive-virtual-storage-views}{https://gitlab.rlp.net/fschuhkn/adaptive-virtual-storage-views}

\vspace*{-0.2cm}
\bibliographystyle{ACM-Reference-Format}
\bibliography{virtual_views}

\end{document}